\documentclass[aps,pra,showpacs,amsmath,amssymb,preprintnumbers,superscriptaddress,10pt,twocolumn,longbibliography]{revtex4-1}

\usepackage{graphicx}
\graphicspath{{figures/}}
\usepackage{SIunits}
\usepackage{hyphenat}
\usepackage[bookmarks=false]{hyperref}
\usepackage{color}
\usepackage[capitalise]{cleveref}
\crefname{section}{Sec.}{Secs.}
\Crefname{section}{Section}{Sections}
\usepackage{physics}

\definecolor{pink}{RGB}{255,0,255}

\definecolor{blue}{RGB}{0,0,205}

\newcommand{\rev}[1]{#1}

\begin{document}

\title{Independent \rev{quality assessment} of a commercial quantum random number generator}

\author{Mikhail~Petrov}
\email{m.petrov@rqc.ru}
\affiliation{Russian Quantum Center, Skolkovo, Moscow 121205, Russia}
\affiliation{NTI Center for Quantum Communications, National University of Science and Technology MISiS, Moscow 119049, Russia}

\author{Igor~Radchenko}
\affiliation{Moscow State University, Moscow, 119991 Russia}

\author{Damian~Steiger}
\affiliation{Institute for Theoretical Physics, ETH Zurich, CH-8093 Zurich, Switzerland}
\affiliation{Microsoft Corporation, One Microsoft Way, Redmond, WA 98052, USA}

\author{Renato~Renner}
\affiliation{Institute for Theoretical Physics, ETH Zurich, CH-8093 Zurich, Switzerland}

\author{Matthias~Troyer}
\affiliation{Institute for Theoretical Physics, ETH Zurich, CH-8093 Zurich, Switzerland}
\affiliation{Microsoft Corporation, One Microsoft Way, Redmond, WA 98052, USA}

\author{Vadim~Makarov}
\affiliation{Russian Quantum Center, Skolkovo, Moscow 121205, Russia}
\affiliation{\mbox{Shanghai Branch, National Laboratory for Physical Sciences at Microscale and CAS Center for Excellence in} \mbox{Quantum Information, University of Science and Technology of China, Shanghai 201315, People's Republic of China}}
\affiliation{NTI Center for Quantum Communications, National University of Science and Technology MISiS, Moscow 119049, Russia}
\affiliation{Department of Physics and Astronomy, University of Waterloo, Waterloo, ON, N2L~3G1 Canada}

\date{\today}

\begin{abstract}
We reverse-engineer, test and analyse hardware and firmware of the commercial quantum-optical random number generator Quantis from ID~Quantique. We show that $>99\%$ of its output data originates in physically random processes: random timing of photon absorption in a semiconductor material, and random growth of avalanche owing to impact ionisation. Under a strong assumption that these processes correspond to a measurement of an initially pure state of the components, our analysis implies the unpredictability of the generated randomness. We have also found minor non-random contributions from imperfections in detector electronics and an internal processing algorithm, specific to this particular device. Our work shows that the design quality of a commercial quantum-optical randomness source can be verified without cooperation of the manufacturer and without access to the engineering documentation. 
\end{abstract}

\maketitle

\section{Introduction}
\label{sec:intro}

Random number generators (RNGs) are used in a large variety of applications. Nowadays both software and physical RNGs are in use \cite{chan2017}. A crucial aspect of random numbers is their unpredictability---the outcome of a coin toss would not be considered random if it could be known before the tossing. Software RNGs do not satisfy this criterion (unless further assumptions are made) for their output is generated by a deterministic algorithm, which is why they are also termed ``pseudo-random''~\cite{knuth1997}.  Conversely, the output of physical RNGs is obtained by measuring physical quantities. According to quantum theory, for a suitably designed measurement on a quantum system, the outcome cannot be predicted even if the system's physical state at the time when the measurement process is started is known completely.  Quantum RNGs exploit such quantum measurements. Hence, if designed properly, their outputs are fundamentally unpredictable and, in this sense, truly random~\cite{herrero2017}.

Although physical RNGs are used in commercial applications, as of yet there does not exist any complete and reliable procedure for their certification \cite{chan2017}. Attempts to establish requirements based exclusively on an analysis of the output stream like NIST's tests~\cite{soto1999} are not sufficient to ascertain randomness, because a statistical test of a sequence can never prove its unpredictability~\cite{rukhin2010}. Indeed, a device may pretend to generate randomness while actually replaying a bit sequence that has been prerecorded from a true random source. The output of such a fake RNG would then obviously pass any statistical test that the true random source passes, whereas a third party may hold an exact copy of the prerecorded sequence and hence predict its output.  

There are in principle two different approaches to resolve this problem. One is device-independent random number generation~\cite{Colbeck_2011,PironioRandomness,ColbeckRandomness}. Here the idea is to consider data generated by two separated devices that share quantum entanglement. The quantum origin of the data can then be certified by a Bell test. The advantage of this approach is that no assumptions about the inner workings of the devices that produce the data are necessary. However, with today's technology, device-independent schemes are complex lab experiments with impractically low bit rates (see, e.g.,\ \cite{liu2021}). In addition, they still need some trusted randomness as input, which is used for selecting between the different observables that enter the Bell test.

\rev{Semi-device-independent QRNG \cite{pawlowski2011} is a more technically feasible approach than device-independent. In return for the relative simplicity of the QRNG implementation and increase generation rate, the semi-device-independent QRNG requires some assumptions about the device operation or its features, although still does not need a complete device model. As example, some semi-device-independent protocols require that QRNG should have trusted source \cite{nie2016,cao2015} or trusted measurement \cite{vallone2014,avesani2018}. Other protocols do not require any assumptions for setup components, but they make assumptions on the overlap \cite{brask2017} or the energy \cite{ruska2019} of the prepared quantum states or assumptions on the Hilbert space dimension \cite{lunghi2015,mironowicz2021}. While generation rate increased significantly, technical realization of the semi-device-independent QRNG remains relatively complex and there are still no on-shelf devices.}

In this work we are concerned with the converse, i.e.,\ the device-dependent, approach~\cite{frauchiger2013}. In contrast to the above, device-dependent RNGs are more practical, smaller, faster, and cheaper~\cite{herrero2017}. The price to pay for this is that, to certify their unpredictability, one requires an accurate and verifiable model of the device's operation, described within the formalism of quantum theory. Such a description is however rarely available for real-world devices. In this case the assessment of the quality of the generated randomness  may still be based on tests of the individual components of the device, but must usually be supplemented by strong assumptions about the parts that cannot be analysed completely. 

Here we carry out such an analysis for the quantum-optical ``Quantis'' device from ID~Quantique \cite{Quantis}, which has been available since the year 2001 and used in a number of real-world applications (Swiss Lottery, United Kingdom NSI Banking, Ukraine Online Gaming, etc.\ \cite{Quantis}). As explained above, the question whether the device generates true randomness cannot be answered by mere statistical tests of the output sequence. Instead, a user must trust that the randomness-generating process the device's manufacturer claims to employ has been implemented correctly. To establish this trust, it should  be possible for an independent party to examine and verify the generator, including an in-depth inspection of its internal functioning. This certification can be commissioned by the manufacturer from an accredited certification lab. ID~Quantique has got verified the compliance of Quantis with the AIS~20/31 standard \cite{walenta2015,Quantis-AIS31}. 

Typically, in the course of such a certification, the lab examines design documentation provided by the manufacturer, and examines a device sample according to procedures defined in the standard. However, it is not clear how well the procedures in the existing standards cover the greater number of physical phenomena in the quantum RNG. Therefore, here we perform an independent examination of Quantis without access to the manufacturer's internal documentation. The goal of our work is to identify the physical processes that produce data in Quantis, and verify that the internal post-processing of this data is sound. Our analysis and model are specific to this particular type of device. Other types of QRNGs would have a different model and, possibly, other analysis approaches of their hardware, operation, and post-processing algorithms.

In  analogue to existing practices in highly-demanding hardware-dependent areas \cite{youn2014,kornecki2010}, the certification procedure of the physical RNG should consists of at least the following four stages.
\begin{itemize}
 \item Discussion of the underlying physical model and assumptions.
 \item Examination of calculation algorithms.
 \item Inspection of the hardware realization.
 \item Statistical tests of the output bit stream.
\end{itemize}

We follow the above methodology in our study. Previous studies have only tested the output stream of Quantis \cite{calude2010,abbott2014,walenta2015,Quantis_S_Test} but not analysed its internals. 

We have examined 6 devices with different manufacturing dates, ranging from 2007 to 2013 with serial numbers (s/n): 0701100A210, 0701108A210, 0701132A210, 0902242A210, 1304527A210, and 1304609A210 (the first two digits represent the year of manufacture and the remaining digits are internal serial numbers). Our key sample that provides most of our data has been purchased from a regular stock, without warning the manufacturer of its intended use. We have been guided only by openly available information: a white paper \cite{Quantis_W_Paper}, application note \cite{Quantis_A_Note}, user guide \cite{Quantis}, randomness test report \cite{Quantis_S_Test}, and a patent that outlines the actual implementation of the optics \cite{ribordy2009}. These sources provide a very basic understanding of the device's principle and functionality. To obtain the rest of the necessary data, we have reverse-engineered the device, examining and analysing closely its electrical and optical parts. During the examination Quantis s/n~0902242A210 has been destroyed in order to explore its optical part, obtain images of avalanche photodiodes (APDs) and measure properties of the light source. Our key sample (s/n~1304527A210) has been disassembled but remained functional, and all our in-vivo measurements have been done on it. The other four samples (s/n~1304609A210, 0701100A210, 0701108A210, and 0701132A210) have not been disassembled and have only been used for tests of their output bit stream.

\section{Quantis teardown}
\label{sec:teardown}

\begin{figure}
	\includegraphics{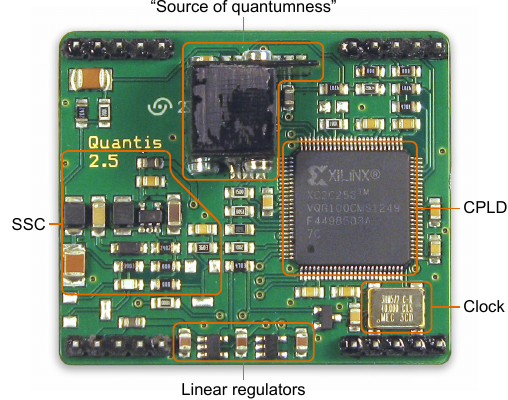}
	\caption{Main PCB, component side. SSC - step-up switching DC/DC converter, CPLD - complex programmable logic device, Clock - system clock.}
	\label{fig:Main_PCB_top}
\end{figure}

The basis of Quantis hardware is a printed circuit board (PCB) that carries all its construction elements. The board is coated with a thick layer of black epoxy then packed into a metal can, presumably either to hide the design or to protect internal components from ambient light and moisture. We have removed the can and epoxy by heating the PCB up to about $150~\celsius$ with a hot-air gun. At this temperature solder does not yet melt and electronic components survive, while the epoxy softens and can be peeled off completely. The PCB is shown in \cref{fig:Main_PCB_top}.

\begin{figure}
	\includegraphics{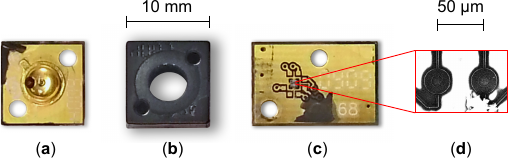}
	\caption{``Source of quantumness'' taken apart. (a)~Light emitting diode (LED) light source. (b)~Anodized aluminum sleeve. (c)~Pair of single-photon detectors. (d)~Photosensitive areas of the single-photon detectors (electron-microscope image).}
	\label{fig:Source_of_quantumness}
\end{figure}

A key part of the device is its ``source of quantumness'', consisting of a black anodized aluminum sleeve [\cref{fig:Source_of_quantumness}(b)] with a light source at one end [\cref{fig:Source_of_quantumness}(a)] and a pair of single-photon detectors at the opposite end [\cref{fig:Source_of_quantumness}(c)]. No optical beamsplitter element has been found inside the sleeve, which is consistent with the patent \cite{ribordy2009} but disagrees with the optical scheme included in the specification of the device that shows a free-space beamsplitter (Fig.~1 in \cite{Quantis_W_Paper}).

The detectors are avalanche photodiodes (APDs) working in a Geiger mode \cite{cova2004}. The all-silicon structure embodies a pair of APDs, amplifiers and quenching circuit for them. Geometric dimensions of the APDs have been determined by electron microscopy: the sensitive areas have a round shape with a diameter of $\approx 10~\micro\meter$ spaced at $50~\micro\meter$ between their centers [\cref{fig:Source_of_quantumness}(d)]. A programmable step-up switching DC/DC converter (SSC) provides a bias voltage for both APDs.

We have measured spectral characteristics of the light source. It has broadband emission centered at $820~\nano\meter$ with full-width at half-magnitude (FWHM) of $40 \text{~nm}$, and viewing angle of $10\degree$. These characteristics are very typical for a near-infrared light-emitting diode (LED).

Linear voltage regulators with $3.3$ and $1.8~\volt$ output voltages (\cref{fig:Main_PCB_top}) power all on-board electronics. A broad-spectrum $40~\mega\hertz$ oscillator provides a system clock. A complex programmable logic device (CPLD) performs most of the device functionality. This CPLD is Xilinx type XC2C256 in a 100-pin package VQG100CMS1249  with multi-voltage input and output operation from $1.5$ to $3.3~\volt$. Unsurprisingly, the CPLD firmware is locked against its readout. All the following knowledge has therefore been obtained by analysing the rest of the electronic circuit, in-vivo signal capturing, applying external probing signals and observing them propagating through the circuit.

\begin{figure}
	\includegraphics{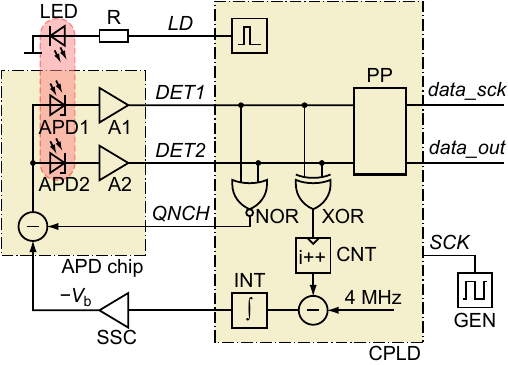}
	\caption{Simplified electrical scheme of Quantis. A, amplifier; APD, avalanche photodiode; CNT, counter; CPLD, complex programmable logic device; GEN, clock generator; INT, integrator; LED, light emitting diode; NOR, inverted OR gate; PP, post-processing algorithm; R, resistor; SSC, switching power supply; XOR, exclusive OR gate.}
	\label{fig:Scheme}
\end{figure}

\begin{figure*}
	\includegraphics{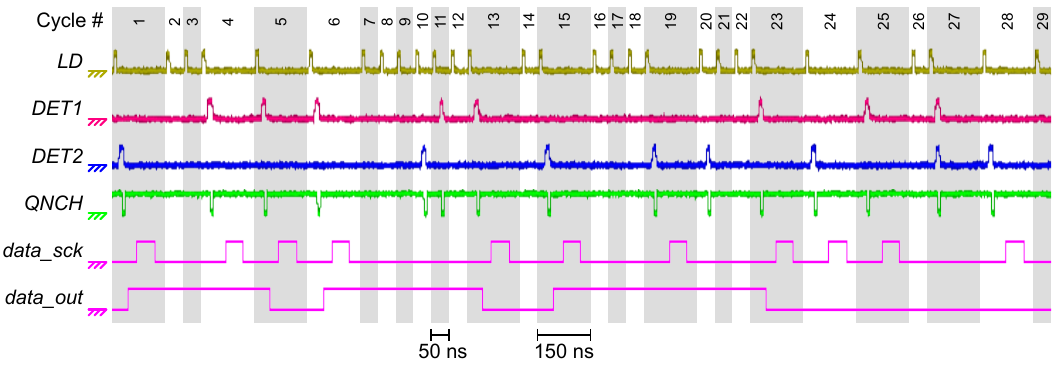}
	\caption{Signals in the circuit, recorded during normal operation. Trace names correspond to signal names in \cref{fig:Scheme}.}
	\label{fig:Captured_signals}
\end{figure*}

\begin{figure}
	\includegraphics{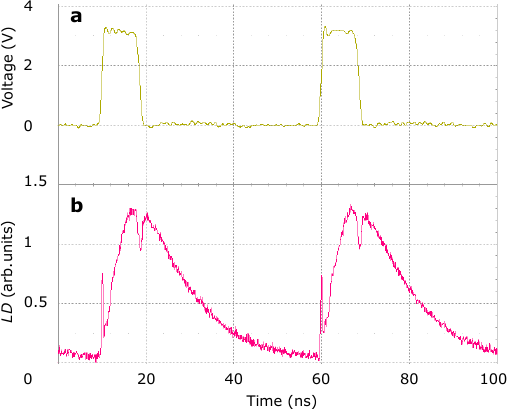}
	\caption{Operation of the light source. (a) Voltage pulses applied to the LED from the CPLD, via line {\it LD}. (b) Light emitted by the LED.}
	\label{fig:LED_pulse_prel}
\end{figure}

\begin{figure}
	\includegraphics{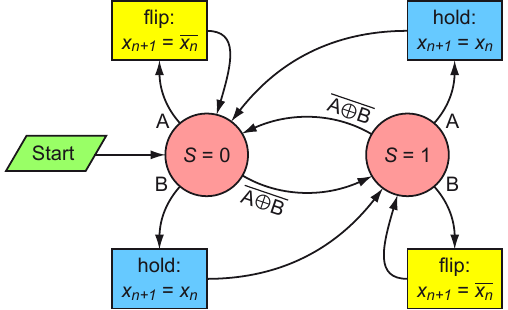}
	\caption{Post-processing state machine.}
	\label{fig:Flowchart}
\end{figure}

A simplified electrical scheme of Quantis and recorded real signals are shown in \cref{fig:Scheme,fig:Captured_signals}. The RNG works in a cycle-based regime. Every cycle starts with a $12~\nano\second$ long voltage pulse formed at line {\it LD} by the CPLD [\cref{fig:LED_pulse_prel}(a)], which is applied through a ballast resistor R to the LED causing it to emit a longer flash of light [\cref{fig:LED_pulse_prel}(b)]. This light may trigger an avalanche in either of the APDs, generating, after amplification, ``click'' signals {\it DET1} and {\it DET2} for the CPLD. The CPLD operation procedure differs depending on whether or not clicks appear from the APDs.  If none of the detectors has clicked, the CPLD just repeat next cycle by the pulse at line LD with a $50~\nano\second$ delay (cycles~\# 2, 3, 7--9, 12, 14,~\dots in \cref{fig:Captured_signals}). If any of the APDs clicks, the CPLD $\textit{DET}x$ input goes high and it activates a quenching procedure by pulling the {\it QNCH} output low, which reduces the bias voltage $V_\text{b}$ on both APDs and thus quenches the avalanche (cycles~\# 1, 4--6, 10, 11, 13,~\dots). In most cycles with detectors clicks (cycles~\# 1, 4--6, 13,~\dots) the CPLD starts next cycle with $150~\nano\second$ delay presumably needed to reduce afterpulsing \cite{cova2004}. But sometimes one or both APDs click with a slight delay relative to the LED light pulse (cycles ~\# 10, 11, 20), in these cases the post-processing algorithm (PP in \cref{fig:Scheme}) does not consider this to be a valid click and the cycle time remains 50 ns. 
 
To convert input APD clicks into the output random bit stream, CPLD performs the post-processing procedure. After post-processing the output binary stream of the RNG is transmitted out through a serial peripheral interface (SPI) bus: the random bit value should be read on {\it \mbox{data\_out}} line at the leading edge of the clock signal {\it \mbox{data\_sck}} \cite{Quantis_A_Note}.  

An analysis of the captured oscillograms reveals the following post-processing algorithm  of converting APD clicks into the output stream. A random bit is output from PP if one and only one APD clicks (cycles \#~1, 4--6, 13, 15, 19, 23--25, 28 in \cref{fig:Captured_signals}), namely the output level {\it \mbox{data\_out}} may change and a sync pulse {\it \mbox{data\_sck}} is generated. In cycles when none (\#~2, 3, 7--9, 12, 14, 16--18, 21, 22, 26, 29) or both (\#~27) APDs click, and in cycles with delayed (\#~10, 11, 20) APD clicks, no output random bit is  produced ({\it \mbox{data\_out}} remains unchanged and there is no sync pulse {\it \mbox{data\_sck}}).
  
The post-processing consist of a state  machine (\cref{fig:Flowchart}). It has two states $S = 0$ and $S = 1$ and generates the output bit $x_n$ ({\it \mbox{data\_out}}) in each CPLD cycle {\it n}. Only one 1-bit internal variable exists: the value $x_n$ of the last random bit outputted (0 or 1). Events $A$ and $B$ correspond to valid clicks of the first ({\it DET1}) and the second ({\it DET2}) APDs, respectively. The state machine works in every cycle as follows. When S = 0 and event A occurs, a ``flip'' is executed---the output bit value is reversed relative to the current one ($x_{n+1}=\overline{x_n}$) and the state S remains unchanged. When S = 0 and event B occurs, a ``hold'' is executed---the value of the output bit does not change ($x_{n+1}=x_n$) and the state S changes to the opposite (S becomes 1). When S = 1, at event A the hold occurs and at event B the flip occurs. In the cases when either none of the events A and B occur or both events A and B occur simultaneously, S changes to the opposite without outputting a bit. Note that PP treats delayed clicks (\#~10, 11, 20) as the absence of A and~B. 
  
\Cref{fig:Captured_signals} shows signals in the circuit well after the state machine has started. For the cycle we numbered \#0 in \cref{fig:Captured_signals} we have S = 1, $x_{0}=0$, and {\it \mbox{data\_out}} = 0.
  
A feedback loop exists to maintain a mean rate of the output stream at the level of $4~\mega\hertz$. For this purpose, the CPLD varies the bias voltage of the APDs $V_\text{b}$, effectively tuning their quantum efficiency (\cref{fig:Scheme}). A counter CNT measures a mean frequency of cycles when only one detector clicks. The error signal of the feedback loop is a difference between the value counted and the target rate of $4~\mega\hertz$. The error signal passes through a software integrator INT and is applied to the voltage control input of the SSC.

\section{Analysis of design}
\label{sec:inspection}

\subsection{Physical model}
\label{sec:P_model}

We now have a closer look at Quantis' underlying physical model, which we describe in terms of standard notions from quantum optics. As we have investigated before, the light source is a LED with central line at $\lambda = 820~\nano\meter$ and bandwidth of $2 \cdot \Delta \lambda = 40~\nano\meter$. For such a source the coherence time (a characteristic period of time while light wave ``remembers'' its history) can be estimated as
\begin{equation}
\tau_\text{coh} = \frac{\lambda^2}{2 \pi c \cdot \Delta \lambda} \simeq 18~\femto\second.
\end{equation}

On the other hand, the registration period of APDs signals is at least $25~\nano\second \gg \tau_\text{coh}$. Hence, the signal measured can be regarded a result of a large number of possible independent photon absorption events. This would imply that the APD clicks have Poissonian statistics and are independent in both channels.

No entanglement by the number of photons may exist in the current scheme: the presence or absence of a photon in one channel says nothing about signal in the other one. It contradicts the principle declared in ID~Quantique Quantis white paper (p.~11) \cite{Quantis_W_Paper} that states it is a ``which way'' scheme. It is not, because registration of a photon in one channel does not exclude the possibility of photon registration in the other channel.

The actual physical source of randomness in Quantis is the photoexcitation of a carrier in the absorption layer of the APD \cite{cova2004}. A secondary significant source of randomness is the subsequent random growth of avalanche by impact ionisation in the APD \cite{spinelli1997}. Owing to the statistical nature of the latter process, some avalanches die without being detected (their number of carriers may fluctuate down to zero), and for those detected their detection time is randomly distributed. 

Lacking a precise microscopic model of this hardware, we cannot however without further assumptions conclude that its apparent random behaviour is due to a generically unpredictable quantum process. At this point we thus need to make a crucial assumption. We suppose that the measured statistics of the data produced by the components would be unchanged if \rev{all degrees of freedom that are accessible to an adversary} were initialised to any pure state. This assumption guarantees that a possible attacker who has access to information about the device's initial state cannot predict its outputs (beyond the bias implied by the measured statistics).

\subsection{Post-processing procedure}
\label{sec:P_processing}

Now, let us consider the post-processing algorithm with assumptions that follow from the physical model. We treat the signals from the pair of APDs as independent Poisson processes with different probabilities of clicks
\begin{equation}
  \mathbb{P}(\mathrm{D_1}) \equiv p_1, ~~~ \mathbb{P}(\mathrm{D_2}) \equiv p_2.
\end{equation}
The probabilities of events when one and only one particular detector clicks (events A and B) are
\begin{equation}
  \begin{array}{l}
  \mathbb{P}(\mathrm{A}) = \mathbb{P} \left ( \mathrm{D_1} \bullet \overline{\mathrm{D_2}} \right ) = p_1 (1 - p_2) \equiv \alpha,\\
  \mathbb{P}(\mathrm{B}) = \mathbb{P} \left ( \mathrm{D_2} \bullet \overline{\mathrm{D_1}} \right ) = p_2 (1 - p_1) \equiv \beta
  \end{array}
\end{equation}
and the probability that neither A nor B takes place is
\begin{equation}
  \mathbb{P} \left ( \overline{\mathrm{A} + \mathrm{B}} \right ) = \mathbb{P} \left ( \overline{\mathrm{D_1} \oplus \mathrm{D_2}} \right ) = 1 - \alpha - \beta.
\end{equation}
In this notation the probability that the next output bit will be inverted with respect to the current bit (we call this action a flip) equals to
\begin{equation}
  \begin{split}
    \mathbb{P}(\mathrm{flip}) = \, & \alpha \left [ \alpha + (1 - \alpha - \beta) \beta \right ] \sum _{m = 0} ^{\infty} (1 - \alpha - \beta) ^ {2m} \\
                                   & + \beta \left [ \beta + (1 - \alpha - \beta) \alpha \right ] \sum _{m = 0} ^{\infty} (1 - \alpha - \beta) ^ {2m} \\
                              = \, & {{\alpha ^2 + \beta ^2 + 2 \alpha \beta (1 - \alpha - \beta)} \over {1 - (1 - \alpha - \beta) ^2}},
  \end{split}
\end{equation}
where the sum over $m$ is the probability of an even number of transitions between $S=0$ and $S=1$ without producing an output bit. Similarly, the probability of the next bit being equal to the current bit (hold) is
\begin{equation}
  \begin{split}
    \mathbb{P}(\mathrm{hold}) = \, & \alpha \left [ \beta + (1 - \alpha - \beta) \alpha \right ] \sum _{m = 0} ^{\infty} (1 - \alpha - \beta) ^ {2m} \\
                                   & + \beta \left [ \alpha + (1 - \alpha - \beta) \beta \right ] \sum _{m = 0} ^{\infty} (1 - \alpha - \beta) ^ {2m} \\
                              = \, & {{2 \alpha \beta + (\alpha ^2 + \beta ^2) (1 - \alpha - \beta)} \over {1 - (1 - \alpha - \beta) ^2}}.
  \end{split}
\end{equation}
Their difference is thus
\begin{equation}
  \begin{split}
    \mathbb{P}(\mathrm{flip}) - \mathbb{P}(\mathrm{hold}) = \, & {{(\alpha - \beta) ^2} \over {2 - \alpha - \beta}} \\
                                                          = \, & {{(p_1 - p_2) ^2} \over {2 - p_1 - p_2 + 2 p_1 p_2}} \ge {{(p_1 - p_2) ^2} \over {2}}.
  \end{split}
\label{eq:XOR_bias_estimation}
\end{equation}
For a real physical system, the count probability of two detectors can never be perfectly equal, owing to differences in their quantum efficiency, size, intensity of illumination, and possibly other factors. It follows from \cref{eq:XOR_bias_estimation} that if the probabilities of APD signals $p_1$  and $p_2$ are not exactly equal, then the event flip will be more likely than hold. This intrinsic property of the PP introduces correlations in the output stream, i.e.,\ makes it less than perfectly random. We have studied this effect experimentally in \cref{sec:Stat_imperfections}.

The prevalence of the flip events may also be caused by the APD signals being non-Poissonian, in particular their exhibiting afterpulsing. We have not considered this effect in our model.

\subsection{Feedback loop stability}
\label{sec:Feedback}

A feedback loop that maintains a constant bitrate of the output stream includes integrator INT and switching power supply SSC (\cref{fig:Scheme}), besides other elements. SSC has a passive filter network at its output with a time constant of $100~\milli\second$. This means that its frequency response decays by $20~\deci\bel$ per frequency decade at frequencies $> 1.6~\hertz$. The integrator provides an additional slope of $20~\deci\bel$ per decade in the loop gain. Hence, the phase margin is not sufficient, which may lead to peaking and oscillation of the output stream bitrate.

Moreover, this feedback loop in theory allows that a lock situation may happen.  With increasing reverse bias voltage $V_\text{b}$, the probability of APD clicks increases. However, only single-detection events are counted. The higher $V_\text{b}$ is, the more simultaneous clicks in both detectors appear and these events will be discarded. The negative feedback may then turn into positive. This may in principle lead to the system locking at the maximum $V_\text{b}$. 

\section{Measurements}
\label{sec:measurements}

In order to test the ability of RNG Quantis to generate random sequences, we have carried out a number of measurements on both the source of randomness (the APDs) and post-processing procedure. The objective of each is to find and quantify possible non-random effects. Since we are measuring correlations at the APD outputs, we deem it unnecessary to check the stability and power of the LED emission. It is clear both APDs are working in the Geiger mode, although their photon detection efficiency may be rather low (we did not measure it). We do not think low photon detection efficiency affects the randomness, because the physics of the avalanche remains the same. We compare the contribution of possible non-random effects in the output bit stream with the specification of Quantis that states that ``thermal noise contribution'' should be less than 1\% \cite{Quantis}, which means to be the upper bound on all potential non-randomness in the output stream. While we cannot claim that our set of measurements is complete, the sum of the non-random effects we have found does not exceed 1\%.

\subsection{Dark counts}
\label{sec:dc}

Even in the absence of light, the APDs produce a certain number of clicks---dark counts \cite{cova2004}. Conservatively, these are not considered to be the source of randomness.

\begin{figure}
	\includegraphics{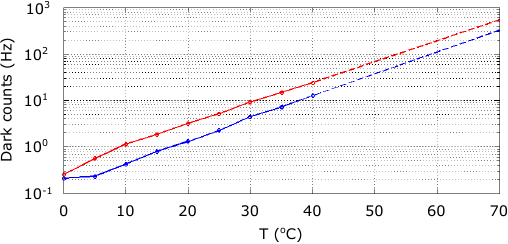}
	\caption{APD dark count rate for the two APDs.}
	\label{fig:APD_dark_counts}
\end{figure}

We have measured the dark counts in $0$ to $40~\celsius$ temperature range, by placing Quantis in a thermal chamber. Results of the measurements are shown in \cref{fig:APD_dark_counts}. The dark count rate rises exponentially with temperature. Extrapolating to $+70~\celsius$, which it a commonly assumed upper end of operating temperature range for commercial products, we obtain less than $1~\kilo\hertz$ dark count rate summed over the two APDs. Thus the dark counts contribute less than $0.025\%$ of the output bits.

\subsection{Autocorrelation of APD counts}
\label{sec:autocorrelation}

In Quantis the sources forming a random output sequence are APDs. Therefore, we have first studied the properties of the output signals obtained from photodetectors directly after their pre-amplification ({\it DET1} and {\it DET2} in \cref{fig:Scheme}).

\begin{figure}
	\includegraphics{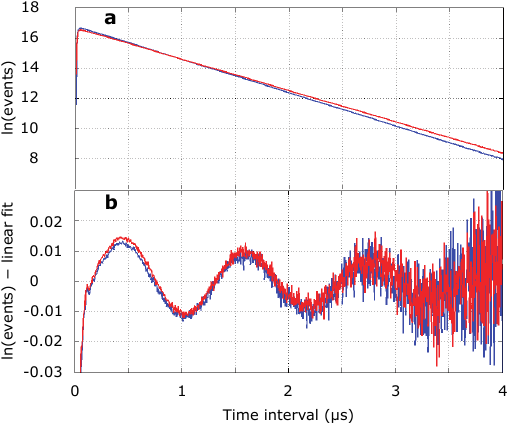}
	\caption{Autocorrelation of APD clicks under continuous-wave illumination, for the two APDs (red and blue). (a)~Plot on the log scale. (b)~Deviations of the measured data from an exponential fit (or linear fit of the log plot). The measurement time was $1000~\second$.}
	\label{fig:APD_wait_time_intervals}
\end{figure}

As discussed in \cref{sec:P_model}, the clicks from each photodetector should be independent and their statistics should be Poissonian. The measured autocorrelation function under continuous-wave illumination from the LED is plotted in \cref{fig:APD_wait_time_intervals}(a). It has an expected shape for a Poissonian process, with a dip in the first $150~\nano\second$ owing to the deadtime imposed by the PP. However a close examination reveals small-amplitude oscillation of unknown origin, which we have magnified in \cref{fig:APD_wait_time_intervals}(b). The peak-to-peak magnitude of these oscillations reaches $2.6\%$.

Owing to the relatively large magnitude exceeding $1\%$ and the oscillation frequency comparable with the output bit rate, this effect is potentially significant. To analyse it, the measurement needs to be repeated in the normal operation of the circuit (with gated LED). Also, a cross-correlation on a similar or longer time scale need to be measured and the combined correlation propagated through the PP. Unfortunately, we have realised this after dismantling the experiment. A simpler cross-correlation measurement presented in \cref{sec:meas-cross-correlation} is insufficient for this analysis.

\subsection{Cross-correlation of APD counts}
\label{sec:meas-cross-correlation}

During an avalanche, the APD emits a few photons, so-called backflash \cite{renker2006}. These may reach the other APD (via internal reflections and scattering inside the optical enclosure shown in \cref{fig:Source_of_quantumness}) and cause a correlated click. Also, electronic interference between the two single-photon detector circuits may in principle exist. Such clicks are not considered to be random.

\begin{figure}
	\includegraphics{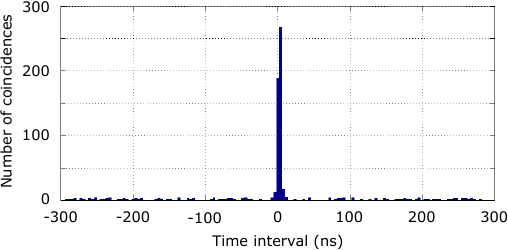}
	\caption{APD cross-talk, measured in darkness over $16~\hour$. Histogram bin size is $4~\nano\second$.}
	\label{fig:APD_crosstalk}
\end{figure}

In order to estimate the click rate owing to the backflash, we have electrically disconnected the LED and measured cross-correlation between {\it DET1} and {\it DET2} in darkness (\cref{fig:APD_crosstalk}). The peak owing to the optical cross-talk is clearly visible. However, the probability of backflash-induced click is small: in $16~\hour$ measurement time, we have registered $2.8 \times 10^6$ single clicks in one APD and $4.3 \times 10^6$ in another, but only about 500 coincidences in $\pm 20~\nano\second$ window. Thus the contribution of the cross-talk to the output bit stream is $\approx 0.007\%$.

\begin{figure}
	\includegraphics{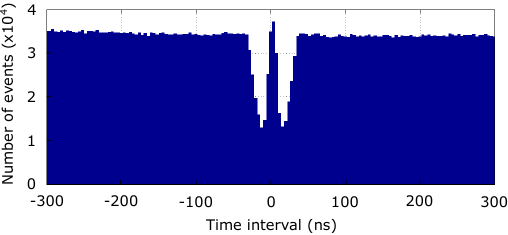}
	\caption{Cross-correlations of APD clicks under continuous-wave illumination, measured over $3600~\second$. Histogram bin size is $4~\nano\second$.}
	\label{fig:APD_crosscorrelations_CW}
\end{figure}

\begin{table*}
	\caption{Output stream statistics. Each sequence length $N = 1 \text{~Gibit} ~ (\equiv 2^{30} \text{~bit})$.}
	\begin{tabular}[t]{*{6}{c@{\quad}}c}
	\hline\hline
	Quantis s/n &
	$\mathbb{N}(x_n = 0)$ & $\mathbb{N}(x_n = 1)$ &
	${\mathbb{N}(= 1) - \mathbb{N}(= 0)} \over {\mathbb{N}(= 1) + \mathbb{N}(= 0)}$ &
	$\mathbb{N}(x_n \oplus x_{n + 1} = 0)$ & $\mathbb{N}(x_n \oplus x_{n + 1} = 1)$ &
	${\mathbb{N}(\oplus = 1) - \mathbb{N}(\oplus = 0)} \over {\mathbb{N}(\oplus = 1) + \mathbb{N}(\oplus = 0)}$ \\
	\hline
	0701100A210 & 536867999 & 536873825 & $~~5.4 \times 10^{-6}$ & 536828388 & 536913435 & $ 7.9 \times 10^{-5}$ \\
	0701108A210 & 536869215 & 536872609 & $~~3.2 \times 10^{-6}$ & 536839365 & 536902458 & $ 5.9 \times 10^{-5}$ \\
	0701132A210 & 536892157 & 536849667 & $-4.0 \times 10^{-5}$ & 536666863 & 537074960 & $ 3.8 \times 10^{-4}$ \\
	1304527A210 & 536882563 & 536859261 & $-2.2 \times 10^{-5}$ & 536787990 & 536953833 & $ 1.5 \times 10^{-4}$ \\
	1304609A210 & 536873035 & 536868789 & $-4.0 \times 10^{-6}$ & 536698339 & 537043484 & $ 3.2 \times 10^{-4}$ \\
	\hline\hline
	\end{tabular}
	\label{tab:Statistic_corruption}
\end{table*}

In order to check for possible further cross-talk effects, we have repeated the measurement under continuous-wave illumination from the LED. The result is shown in \cref{fig:APD_crosscorrelations_CW}. The central features are caused by the expected circuit operation such as quenching (\cref{sec:teardown}). However any cross-correlation beyond the shortest bit generation interval of $\pm 150~\nano\second$ would be of interest, because it may affect the output bit stream. Our histogram shows an elevated cross-correlation probability in $-300$ to $-150~\nano\second$ range, however a further study is required to confirm and quantify it.

\subsection{Statistical imperfections in the output bit stream}
\label{sec:Stat_imperfections}

To verify statistic bias owing to APD efficiency mismatch derived in \cref{sec:P_processing}, a dedicated FPGA-based circuit has been designed. It allows to analyse long consistent sequences in real time without missing a bit. With this setup, the sequence of signals of $N = 1$~Gibit length at the output of the RNG has been analysed. The results of this analysis are given in \cref{tab:Statistic_corruption}. We have measured five devices with different s/n. For each device, we have counted the number of bits 0 and 1 in the output stream $\mathbb{N}$ and calculated their relative deviation from equiprobable. We have also counted the number of hold and flip events, i.e.,\ the number of two consecutive bits having matching and not matching values. The last column shows the relative deviation of hold and flip from equiprobable.

For a large and perfectly random binary sequence, the standard deviation of the relative deviation from equiprobable is $N^{-1/2} \approx 3.05\cdot 10^{-5}$. The relative deviation between the number of 0 and 1 bits in our tests is small and does not exceed the standard deviation, with the exception of the device s/n~0701132A210 that slightly exceeds it. These results are in good agreement with the expected equal probability of 0 and 1, i.e.,\ the output sequence is balanced. However, for the hold and flip events the situation is different. Their measured relative deviation exceeds the standard deviation by a factor of 2 to 12. We infer that this statistical deviation is due to APD efficiency mismatch. 

Assuming click probabilities for both APDs are approximately equal $p_1 \approx p_2 \approx 0.28$ (estimated from the recorded oscillograms in \cref{fig:Captured_signals}), we obtain from \cref{eq:XOR_bias_estimation}:
\begin{eqnarray}
\label{eq:XOR-bias-estimation-exp}
  \mathbb{P}(\mathrm{flip}) - \mathbb{P}(\mathrm{hold}) \approx &&\; {{(p_1 - p_2) ^2} \over {2 - 2 p_1 + 2 (p_1)^2}} \approx {{(p_1 - p_2) ^2} \over {1.6}},\nonumber\\
  \abs{p_1 - p_2} \approx &&\; \sqrt{1.6 \left[\mathbb{P}(\mathrm{flip}) - \mathbb{P}(\mathrm{hold})\right]}.
\end{eqnarray}
For Quantis s/n~0701132A210, in which the greatest deviation has been observed, the absolute difference of APD click probabilities $\abs{p_1 - p_2} \approx 0.025$ and the relative difference ${\abs{p_1 - p_2} / p_1} \approx 8.8\%$. While this is a fairly good click rate matching for the APDs manufactured on the same chip, they are not identical.

We remark that the above statistically significant prevalence of the flip bit pairs over hold bit pairs has neither been detected by the manufacturer's statistical testing \cite{Quantis_S_Test} nor our own application of the NIST SP800-22 test suite \cite{NIST_SP800-22} on the output stream from our above-mentioned worst sample. It was detected by independent researchers \cite{abbott2014}, who however could not explain its origin. They tested a Quantis sample purchased in 2004 and also observed a statistically significant bias (fewer zeros than ones in the output sequence) that we did not observe in our samples.

\subsection{Feedback signal}
\label{sec:Feedback_signal}

\begin{figure}
	\includegraphics{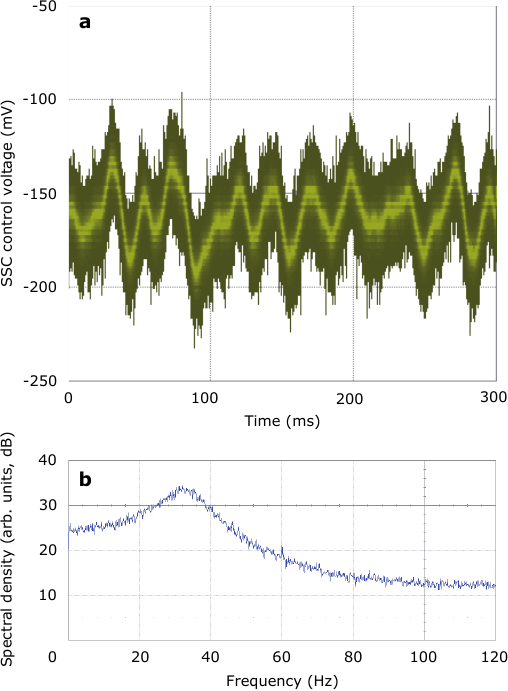}
	\caption{Control input signal of SSC, in (a)~time domain and (b)~frequency domain.}
	\label{fig:Rate_feedback_response}
\end{figure}

We have measured time and frequency characteristics of the feedback signal (\cref{fig:Rate_feedback_response}). As expected, it exhibits oscillations with the spectral maximum around $33~\hertz$. These oscillations however should not affect the probability distribution of the output sequence, because they affect both APDs in the same way. However, they should cause the timing of the output bits to not be regular, which indicates that the timing should not be used in an application.

We have not observed the system locking, whose theoretical possibility is mentioned in \cref{sec:Feedback}.

\section{Discussion and Conclusion}
\label{sec:conclusion}

While no optical beamsplitter element has been found in the Quantis device, it nevertheless contains two  sources of randomness---two Geiger-mode APDs.  Within them, the relevant quantum processes are photoexcitation and impact ionisation. Basically, either APD may be regarded as an independent source of randomness, however the presence of two of them increases the output bit rate. Indeed, a similar QRNG based on a single APD can be constructed \cite{radchenko2015}.

To assess the quality of the randomness generated by these APDs, one would in principle need a microscopic model describing their workings. Within such a model, one may then attempt to prove that their output is unpredictable even if the quantum state of the APDs was fully known (i.e.,\ pure) at the time when the randomness generation process is initiated, that is, when the device received the trigger signal requesting it to  generate randomness~\cite{frauchiger2013}. However, lacking such a microscopic model, one may also resort to physically reasonable assumptions. Specifically, we assume here that the experimentally measured behaviour of the APDs is identical to the one they would exhibit if their microscopic degrees of freedom \rev{that are accessible to an adversary} were at the beginning of each measurement in a pure state. \rev{Under the assumption that the adversary has no access to the device, this assumption holds trivially.}

We have tested for potential imperfections in Quantis that could have an impact on the randomness in the output bit stream. We have found a correlation between adjacent output bits owing to the click rate mismatch of the APDs. However this and other effects stay well below the specified ``thermal noise contribution'' of less than 1\% \cite{Quantis}. Our preliminary conclusion is that Quantis conforms to its published specification of the physical randomness content in the output bit stream, provided that one is ready to make the (strong) assumptions described above. 

Unfortunately, one potential effect that may lead to an additional reduction of randomness---auto- and cross-correlations of APD clicks---has not been sufficiently well measured and analysed by us to reach a conclusion. This could be the topic of a future study. 

We also note that the post-processing implemented by the device does not include randomness extraction. The generated randomness may thus be used for applications where a small bias is acceptable. However, for applications that require uniform randomness, the raw randomness generated by the device would need to be further processed by randomness extractors (see~\cite{Quantis-rand-extractor-techpaper} for details). To choose the corresponding extractor parameters, one would also need an estimate of the (smooth) min-entropy of the raw randomness. Such an estimate would however require additional assumptions on the type of side information held by an adversary as well as a detailed analysis of cross-correlations, and thus goes beyond the scope of this work. \rev{We remark that not all applications require or indeed can tolerate the time-consuming randomness extraction. An example of the latter is testing for the violation of a Bell inequality with the locality and freedom-of-choice loopholes closed \cite{scheidl2010,hensen2015a,hu2018}. There, the short time between the photon absorption in the APD and the resulting random bit being used for measurement choice is a crucial experimental and conceptual constraint.}

Overall, we have shown that an independent security analysis of a commercial quantum RNG can be done. This improves the trust in these devices.

We shared the finished manuscript with ID~Quantique before its submission for publication. The company read it, thanked us, and did not suggest any significant corrections.

\bigskip

\noindent \textbf{Acknowledgements}\\
We thank D.~Frauchiger for discussions.

\medskip

\noindent \textbf{Funding}\\
This work was supported by Industry Canada, CFI, NSERC, Ontario MRIS, U.S.\ Office of Naval Research, Ministry of Education and Science of Russia (program NTI center for quantum communications), and Russian Science Foundation (grant 21-42-00040).

\medskip

\noindent \textbf{Availability of data and materials}\\
Raw experimental data and calculations can be obtained from the corresponding author upon a reasonable request.

\medskip

\noindent \textbf{Competing interests}\\
The authors declare that they have no competing interests.

\medskip

\noindent \textbf{Authors' contributions}\\
M.P.\ finished the data analysis and finished writing the Article with input from all authors. I.R.\ performed all the experiments, hardware, and data analysis, and started writing the Article with input from all authors. D.S.\ and R.R.\ contributed to the data analysis. M.T.\ performed the initial hardware analysis and supervised the study. V.M.\ supervised the study.

\def\bibsection{\medskip\begin{center}\rule{0.5\columnwidth}{.8pt}\end{center}\medskip} 
\bibliography{library}

\end{document}